\documentclass[amsmath,preprintnumbers,10pt,twocolumn,prl]{revtex4-1}
\usepackage{amsbsy}
\usepackage{graphicx}
\usepackage{color}
\usepackage{subfigure}

\usepackage{verbatim}

\begin{document}
\title{Topologically protected modes in non-equilibrium stochastic systems}
\author{Arvind Murugan$^{1,2}$ and Suriyanarayanan Vaikuntanathan$^{*1,3}$} 
\affiliation{${}^1$ James Franck Institute, University of Chicago, Chicago, IL, 60637}
\affiliation{${}^2$ Department of Physics, University of Chicago, Chicago, IL, 60637}
\affiliation{${}^3$ Department of Chemisty, University of Chicago, Chicago, IL, 60637}

\begin{abstract}

Non-equilibrium driving of biochemical reactions is believed to enable their robust functioning despite the presence of thermal fluctuations and other sources of disorder.  Such robust functions include sensory adaptation, enhanced enyzmatic specificity and maintenance of coherent oscillations. Non-equilibrium biochemical reactions can be modeled as a master equation whose rate constants break detailed balance. We find that non equilibrium fluxes can support topologically protected boundary modes that resemble similar modes in electronic and mechanical systems.  We show that when a biochemical network can be decomposed into two ordered bulks that meet at a possibly disordered interferace,  the ordered bulks can be each associated with a topologically invariant winding number. If the winding numbers are mismatched, we are guaranteed that the steady state distribution is localized at the interface between the bulks, even in the presence of strong disorder in reaction rates. We argue that our work provides a framework for how biochemical systems can use non equilibrium driving to achieve robust function. 

\end{abstract}
\maketitle 

Mechanisms used by biological systems to process information and achieve ordered states are far-from-equilibrium and require energy dissipation~\cite{Qian2007}. Kinetic proofreading mechanisms used by the cell to ensure high fidelity copying of genetic material use futile energy consuming cycles to decrease the error rates in DNA replication~\cite{Hopfield1974}. Non-equilibrium forces have also been implicated in the functioning of biochemical networks responsible for adaptation~\cite{Lan2012}, ultra-sensitivity\cite{Tu2013}\cite{Tu2008}, and timing of events in cell cycle~\cite{Cao2015}. 

While the behavior and characteristics of equilibrium systems - where no energy is dissipated - are well known, general principles governing the steady state or fluctuations in it in far-from-equilibrium conditions are just being discovered~\cite{Jarzynski2011}. Here, we show that non-equilibrium stochastic systems can support localized modes that have a character similar to topologically protected boundary modes in mechanical and electronic systems\cite{Kane2013,Hasan2010,Ryu2002,paulose2015topological}. 
Our results are applicable to idealized non-equilibrium Markov state models of kinetic proofreading networks\cite{Murugan2012}, sensory adaptation networks~\cite{Lan2012}, and other biophysical processes. Our results provide a framework for understanding how non-equilibrium biochemical networks are able to perform their function robustly in fluctuating conditions~\cite{Barkai1997}. Non-equilibrium Markov state networks that support topologically protected modes have properties that are much more robust to disorder in kinetic rates, unlike networks without protection.

\begin{figure*}[tbp]
\includegraphics[width=1\linewidth]{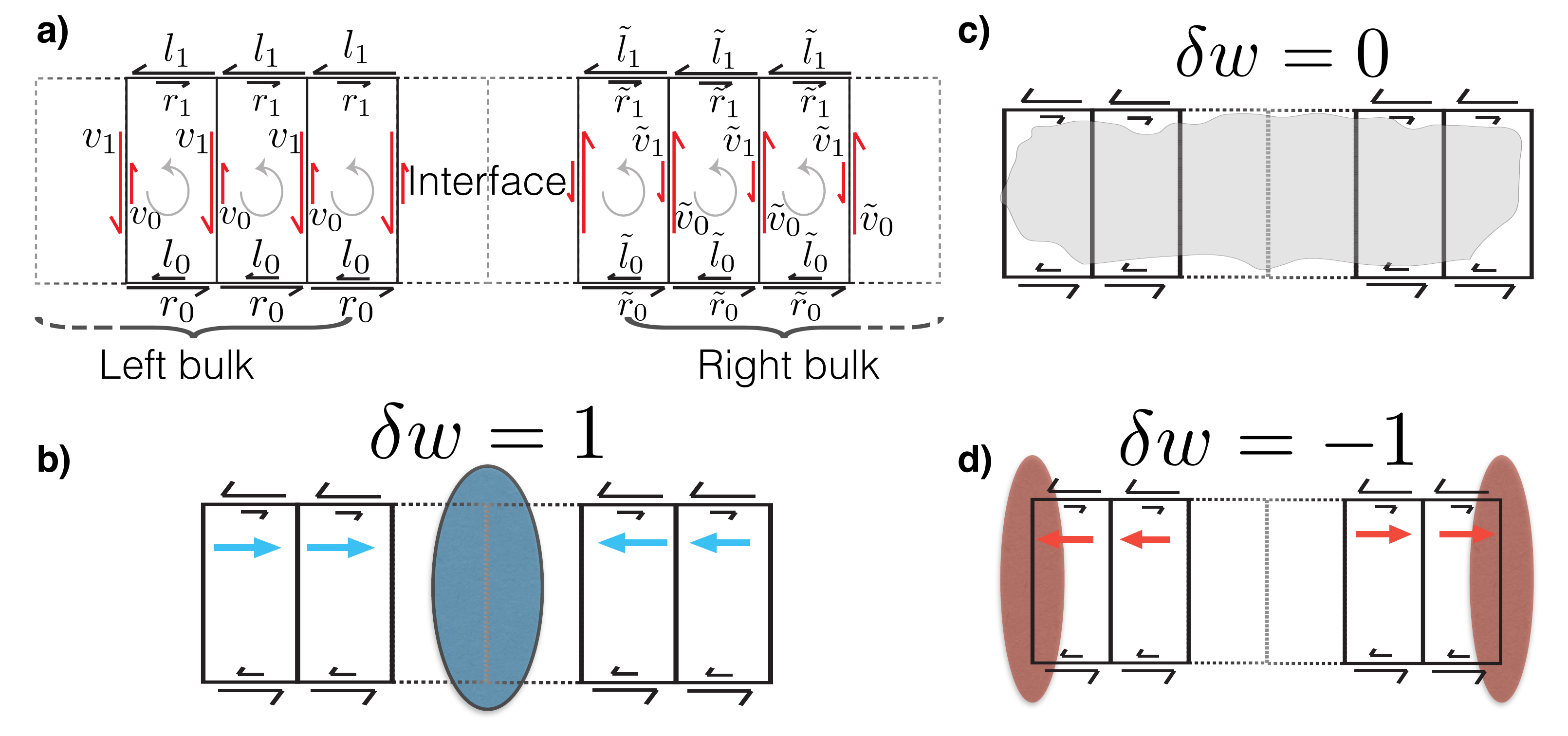}
\caption{(a) A Markov state network with two translationally invariant periodic bulk regions connected together by an interfacial region. Analogous to edge modes in topological insulators and meta-materials, we show that the steady state probability distribution of these Markov state networks is determined by topological winding numbers, defined in Eq.~\ref{eq:labelwinding}, assigned to the bulk domains. 
(b) and (d) The steady state distribution is localized at the boundaries of the bulk domains when the winding numbers of the two bulk phases are mismatched, $\delta w\neq 0 $. We call such states topologically protected states. (c) The steady state is not protected when $\delta w =0$  }
\label{fig:Ladder1}
\end{figure*}

\section{Topological protection and localization in Markov state networks}
We derive our central results in the context of the Markov state model in Fig.~\ref{fig:Ladder1} (a). The Markov state model is composed of two translationally invariant \textit{bulk} like regions with an \textit{interface} connecting them. Specifically, the rates of transitions in the \textit{bulk} regions do not depend on the position along the horizontal axis (Fig.~\ref{fig:Ladder1}). The rates in the interfacial region interpolate between the two bulks. The spatial connectivity and structure of this Markov state network was motivated by that of networks routinely used to study adaptation~\cite{Lan2012}, kinetic proofreading~\cite{Murugan2012}, and cell signal sensing~\cite{Mehta2012}. 

The dynamics of this network can be modeled using a master equation, 
\begin{equation}
\label{eq:master}
\frac{\partial p}{\partial t}=W p 
\end{equation}
where the vector $p$ contains the probability of occupancy of various nodes in the network and $W$ is a state to state transition matrix.

To establish our central results , we find it convenient to consider the statistics of probability current along the horizontal axis in the network in Fig.~\ref{fig:Ladder1}. 
For that, we construct the closely related \textit{tilted} current matrix $W(\lambda)$ with elements
\begin{equation}
\label{eq:tilted}
W(\lambda)_{i,j}=W_{i,j} e^{\lambda (i_x-j_x)}
\end{equation}
where $i_x$ denotes the location of the node $i$ along the horizontal axis. The largest 
eigenvalue of $W(\lambda)$, $\rm{e}(\lambda)$, is the cumulant generating function 
for currents along the horizontal axis, $J$ in the network~\cite{Lebowitz1999}\cite{Garrahan2007}. In particular, $\frac{d{\rm e}(\lambda)}{d\lambda}|_{\lambda=0} =- \langle J\rangle$, giving the net average macroscopic current.
To ensure the possibility of a nonzero current along the horizontal axis, we assume periodic boundary conditions and link up the left and right \textit{bulk} networks through a second interface for our theoretical analysis. We will use $W_{L/R}(\lambda)$ to denote the tilted current matrix of the left (L) and right (R) bulk regions.

{We are interested in conditions under which the steady state probability specified by Eq.~\ref{eq:master} is localized at the interface between the two bulk regions.  Surprisingly, we will find that 
 that localization of the steady state probability at the interface of the spatially heterogenous network can be predicted by assigning particular topological numbers to the \textit{bulk} networks. The details of the interface are not relevant to the existence of these localized modes. The steady state behavior is simply determined by the topological numbers.  

Translational symmetry in the bulk regions makes it convenient to study their properties in terms of Fourier transforms of $W_{\rm{L/R}}(\lambda)$ (SI). A topological characterization of the bulk region can be obtained by first computing the determinant, $\rm{D}(k,\lambda)$, of the Fourier transformed matrix, where $0<k<2 \pi$ denotes the wave vector, and the determinant ${\rm D}(k,\lambda)=|{\rm D}(k,\lambda)| \exp(\rm{i} \theta (k,\lambda))$ is a complex number with phase $\theta(k,\lambda)$. The determinant is periodic in $k$, ${\rm D}(k+2\pi,\lambda)={\rm D}(k,\lambda)$, by construction. A topological number can be assigned to the bulk network by determining the \textit{winding number} $w$ of the phase $\theta(k,\lambda)$~\cite{Ryu2002}. 
\begin{equation}
\label{eq:labelwinding}
w=\frac{\theta_{}(k+2\pi,\lambda)-\theta_{}(k,\lambda)}{2 \pi}
\end{equation}
Specifically, the \textit{winding number} is one when $\theta(k,\lambda)=\theta(k+2\pi,\lambda) + 2\pi$ and zero when $\theta(k,\lambda)=\theta(k+2\pi,\lambda)$. We will show that the steady state probability distribution is localized when the bulk {winding numbers} are mismatched, $\delta w\equiv w_L-w_R\neq 0$ (Fig.~\ref{fig:Ladder1}). This remarkable topological constraint for the localization of the steady state probability at the interface constitutes our main theoretical result.}

\begin{figure}[tbp]
                \includegraphics[width=0.45\textwidth]{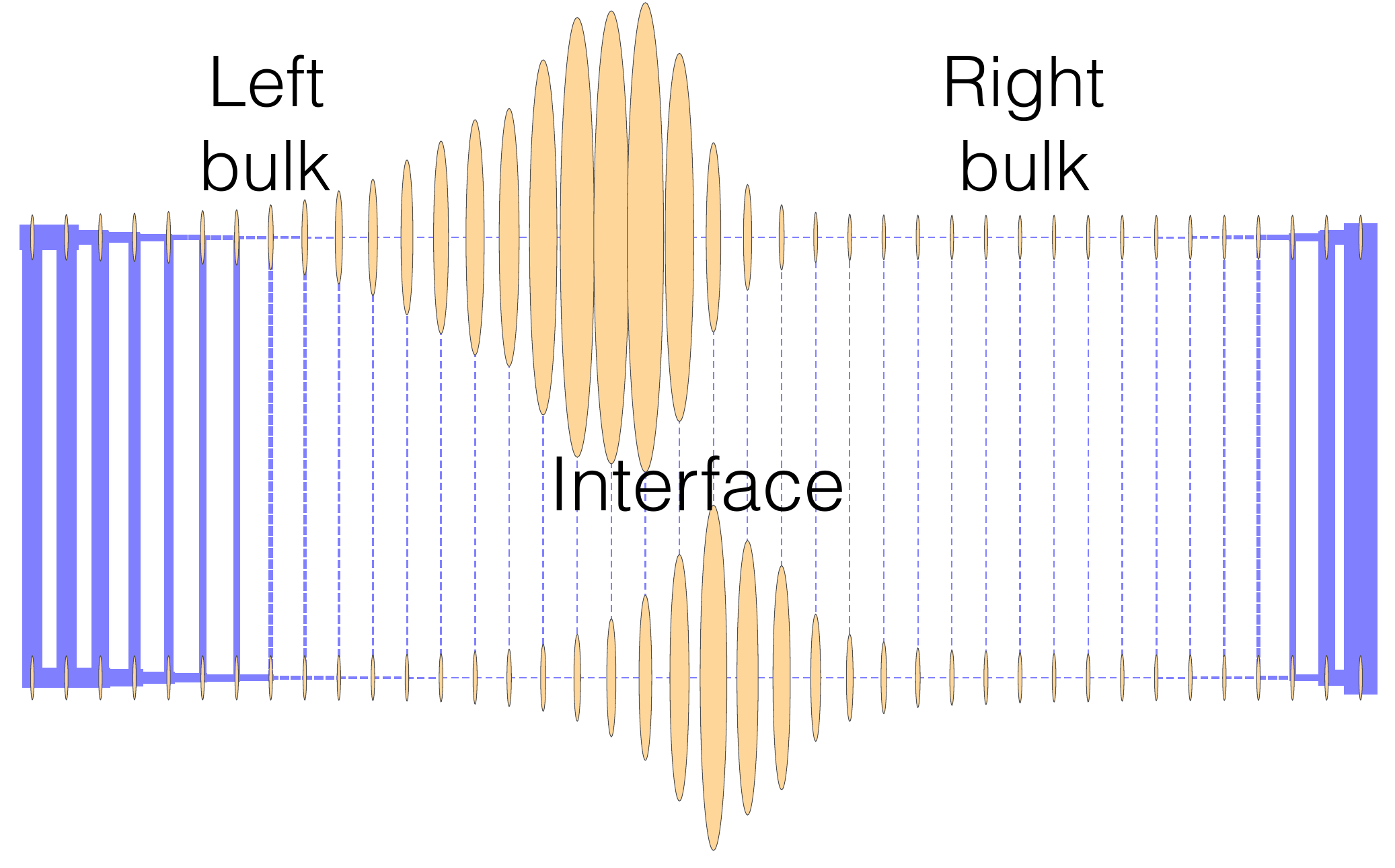}
                 \caption { Numerical results from the ladder network with $\lambda^-<\lambda<\lambda^+$ and $\delta w =1$. Node weights (orange) are obtained from the largest eigenvector of $W(\lambda)$ and link weights (blue) in proportion to elements of the largest eigenvector of $W^{\rm T}(\lambda)$.}
                 \label{fig:Adapt2}
\end{figure}

\subsection{A topological count for the number of localized modes} 

To establish the presence of localized modes in the master equation we will first show that the number of eigenvectors of $W(\lambda)$ with an eigenvalue of zero, henceforth referred to as zero-modes, that are localized at 
the boundary between the left and right \textit{bulk} networks is related to topological invariants computed in the bulk regions. Such 
zero-modes of $W(\lambda)$ imply a zero current along the horizontal axis ($J=0$) and consequently that the probability $p$ in the 
master equation is localized. 

To understand the topological nature of zero-modes of $W(\lambda$) at the boundary between distinct phases, we consider its local index~\cite{Kane2013},
\begin{equation}
\label{eq:indexdefine}
\mbox{ind } W(\lambda) \equiv \mbox{dim} \mbox{ ker} \rho W(\lambda) -  \mbox{dim} \mbox{ ker} \rho W^{\rm T}(\lambda) 
\end{equation} 
where the matrix diagonal matrix $\rho$ has non zero elements $\rho_{i,i}=1$ for nodes $i$ in the interface boundary region~\cite{Kane2013} (Fig.~\ref{fig:Ladder1}), and $\mbox{dim} \mbox{ ker} \rho W(\lambda)$ denotes the dimensionality of the local kernel of $W(\lambda)$. $\mbox{dim} \mbox{ ker} \rho W(\lambda)$ is non-zero if the matrix $W(\lambda)$ has atleast one zero eigen-mode contained in the interfacial region defined by $\rho$. 

While the kernel of $W(\lambda)$ directly corresponds to zero-modes of interest, the zero-modes of $W^{\rm T}(\lambda)$ also have physical significance. Specifically, the elements of these zero-modes, $f_i$, are equal to \begin{equation}
\label{eq:conjugate}
f_i=\lim_{\tau\rightarrow \infty} \langle \exp(-\lambda J)\rangle_i
\end{equation}
where $\langle\dots\rangle_i$ is the average of trajectories, over a long time $\tau$, evolving according to Eq.~\ref{eq:master} conditioned on them beginning at $i$~\cite{Lebowitz1999}. 

As shown in the SI, the index of $W(\lambda)$ can be expressed in terms of topological properties. Specifically, for a system with two bulk regions (as in Fig.~\ref{fig:Ladder1}),
we find that ${\rm ind}\,W$ is given by the difference of two numbers $w_L$ and $w_R$ computed in the left and right bulk phases of the network respectively,
\begin{equation}
\label{eq:widingdifference}
{\rm ind}\,W=\delta w\equiv w_L-w_R
\end{equation}
where,
\begin{equation}
\label{eq:windingdefine}
w_{L/R} \equiv \frac{1}{ 2 \pi \rm{i}} \int_0^{2 \pi} \rm{dk} \partial_k \ln\left[\det [W_{L/R}(\lambda,k)] \right]
\end{equation}
and $W_{L/R}(\lambda,k)$ is the Fourier transform of the tilted translationally symmetric bulk transition matrix (SI). 
The determinant $\det [W_{L/R}(\lambda,k)]$ maps the Fourier transformed matrix to the complex plane, $\det [W_{L/R}(\lambda,k)]=|\det [W_{L/R}(\lambda,k)]| \exp(\rm{i} \theta(\lambda,k))$. The numbers $w_{L/R}$ in Eq.~\ref{eq:windingdefine} hence simply compute the winding number of the phase $\theta(\lambda,k)$ as $k$ is varied from $0$ to $2\pi$. The integers $w_{L/R}$ are the aforementioned \textit{winding numbers}.

\begin{figure}[tbp]
                \includegraphics[width=0.45\textwidth]{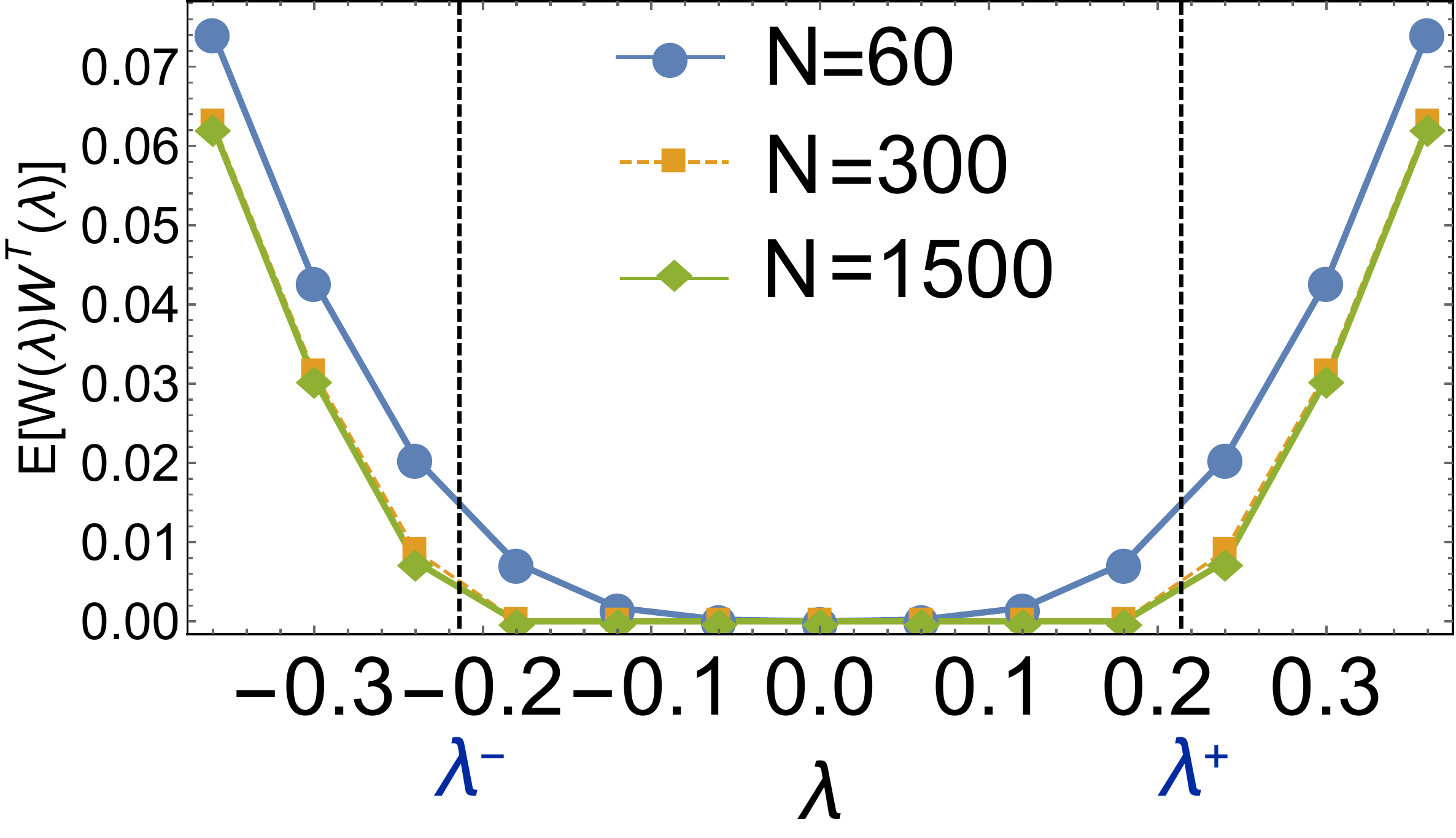}
                 \caption { Lowest eigenvalue (by magnitude) of the operator $W(\lambda)W^{\rm T}(\lambda)$ as a function of $\lambda$. The dotted lines are the theoretical estimate for the region $\lambda^-\leq \lambda\leq \lambda^+$ within which $W(\lambda)W^{\rm T}(\lambda)$, and hence $W(\lambda)$, has a zero eigenmode. The agreement between numerical and theoretical results improves as a function of system size. }
                 \label{fig:Adapt1}
\end{figure}

The above equations predict the existence of \textit{boundary} zero-modes based on the spectrum of $W(\lambda)$ in the \textit{bulk} alone; when $\delta w \neq 0 $, $W(\lambda)$ must have a zero-mode. In the SI we show that if, in fact, $\delta w \neq 0$ for an interval of $\lambda$ around $\lambda = 0$, then the highest eigenvalue of $W(\lambda)$, $e(\lambda)$ is constrained to be zero and the steady state probability distribution of the master equation is localized at the interface. 
 Thus the edge modes of $W(\lambda$) can be simply predicted by computing the \textit{winding} numbers in the bulk of the networks. 

To be concrete, we now numerically demonstrate these behaviors. In Fig.~\ref{fig:Adapt2} we highlight the steady state probability distribution (ellipses) and the steady state distribution of its conjugate defined in Eq.~\ref{eq:conjugate} for parameters that ensure a winding number mismatch of $\delta w=1$. In accordance with the theoretical predictions, the steady state probability distribution is localized at the interface between the two networks, while its conjugate defined by Eq.~\ref{eq:conjugate} is localized away from this interfacial region.

In Fig.~\ref{fig:Adapt1} we identify values of $\lambda$ for which $W(\lambda)$ has a zero eigenvalue for networks of various sizes. The numerical bounds $\lambda^-$ and $\lambda^+$ for which $e(\lambda)$ is constrained to zero agrees with the theoretical predictions obtained from the winding number analysis. The agreement is even more remarkable given that the numerical results were obtained from networks with quenched disordered.  The topological connection allows us to predict the fluctuations of large complex non-equilibrium networks by performing a simple calculation in the bulk regions.  The existence of the points $\lambda^+$ and $\lambda^-$ has further physical 
significance. Since the parameter $\lambda$ is coupled to the rates of transition along the horizontal axis,  $\lambda^+ - \lambda^-$ is related to the effective localization lengths for the probability distributions $\eta$ (see SI),
\begin{equation}
\eta\sim1/(\lambda^+ - \lambda^-)\,.
\end{equation}

The presence of topologically protected edge modes in other contexts is signaled by a gap between the zero energy state and the rest of the energy spectrum of the Hamiltonian operator~\cite{Kane2013}~\cite{Hasan2010}. We numerically observe a similar connection for our dynamical matrix $W(\lambda)$ (SI). The band gap between the maximal and next largest eigenvalue closes as we transition to the unprotected regime. The band gap in the topologically nontrivial regime imposes a constraint on the time scales required to perturb the topologically protected mode and provides another basis for robustness.

\begin{figure}[tbp]
\includegraphics[width=0.5\textwidth]{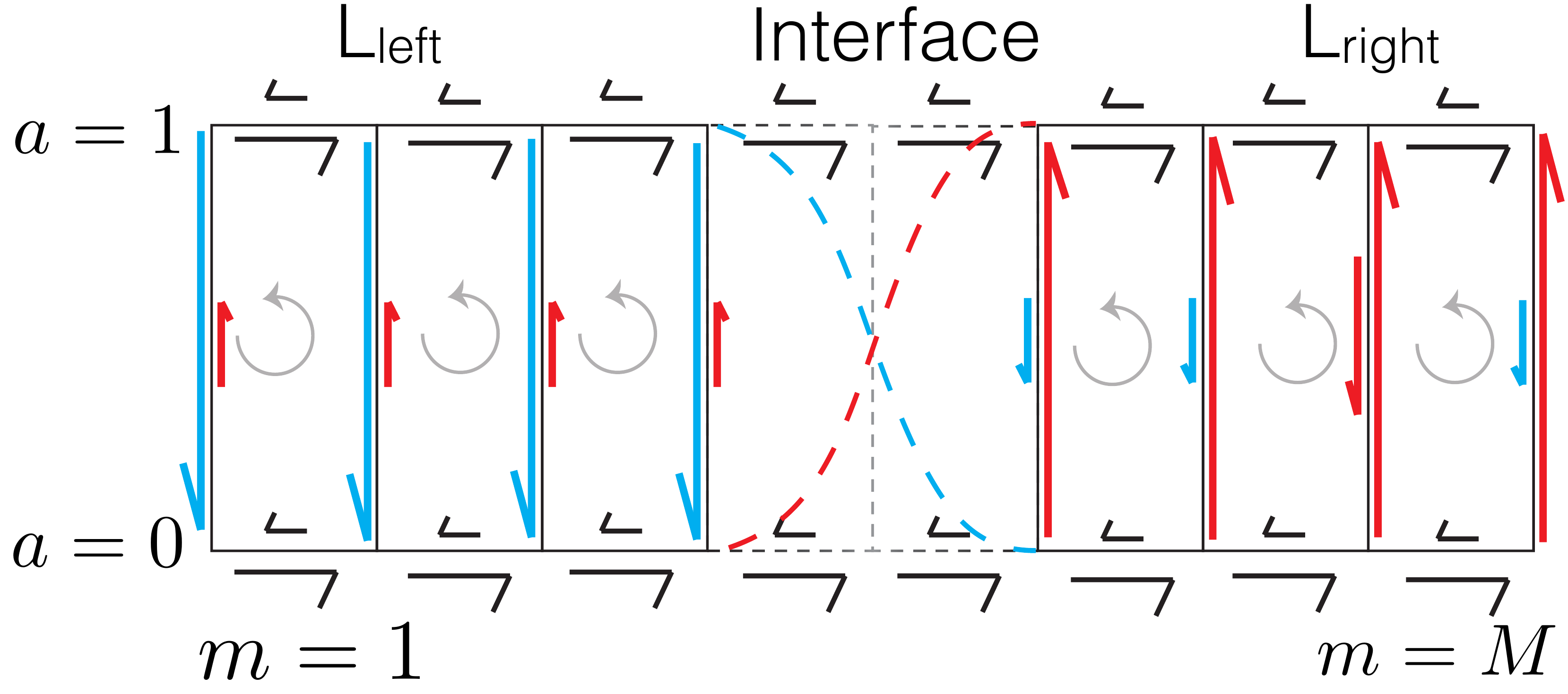}
                 \caption{Markov state representation of an idealized biophysical model of chemosensory adaptation. The rates of transition between the active ($a=1$) and inactive ($a=0$) states are sigmoid functions of the methylation levels ($m$). The transition region of the sigmoid profiles is set by the chemoattractant concentration. The rates of transitions between the various methylation levels are assumed to be independent of the methylation level or chemoattractant concentration in this class of biophysical models. The minimal biophysical model can be viewed a combination of two bulks with an interface between them. }
                          \label{fig:Adapt_example}
                          \end{figure}
               
 \section{Localization and robustness in biophysical networks}
Chemosensory adaptation, kinetic proofreading and many other information processing mechanisms in biology operate far from equilibrium~\cite{Lan2012,Mehta2012}. Topologically protected modes in such networks can enable robust functioning in the presence disorder in the kinetic rates of the network.
We first consider a commonly used idealized biophysical model~\cite{sartori2015free} for chemotaxis adaptation in E.coli. The dynamics of chemotaxis adaptation in E. coli can be described by specifying the activity $a$, and methylation level $m$ of the concentration sensing complex of proteins~\cite{Barkai1997}. 
Transitions between the various mesoscopic states of the protein complex are governed by the Markov state model described in Fig.~\ref{fig:Adapt_example}. 

As discussed in the SI (methods), the transition rates along the activity axis generically are sigmoid functions of the methylation level $m$. The crossover region of these sigmoid functions is set by the chemoattractant concentration sensed by the protein complex. Further, in this class of idealized models, the rates of transition along the methylation axis are independent of methylation level and chemoattractant concentration~\cite{Barkai1997}. This generic sigmoidal profile for the rates of transitions along the activity axis, and the methylation level independent kinetics along the methylation axis establishes the similarity between the chemotaxis adaptation network in Fig.~\ref{fig:Adapt_example} and the network we constructed in Fig.~\ref{fig:Ladder1} (a). The minimal biophysical model for adaptation can hence be viewed as a combination of two periodic bulk networks with an interface between them. The location of the interface is set by the chemoattractant concentration.

\begin{figure}
   \includegraphics[width=0.5\textwidth]{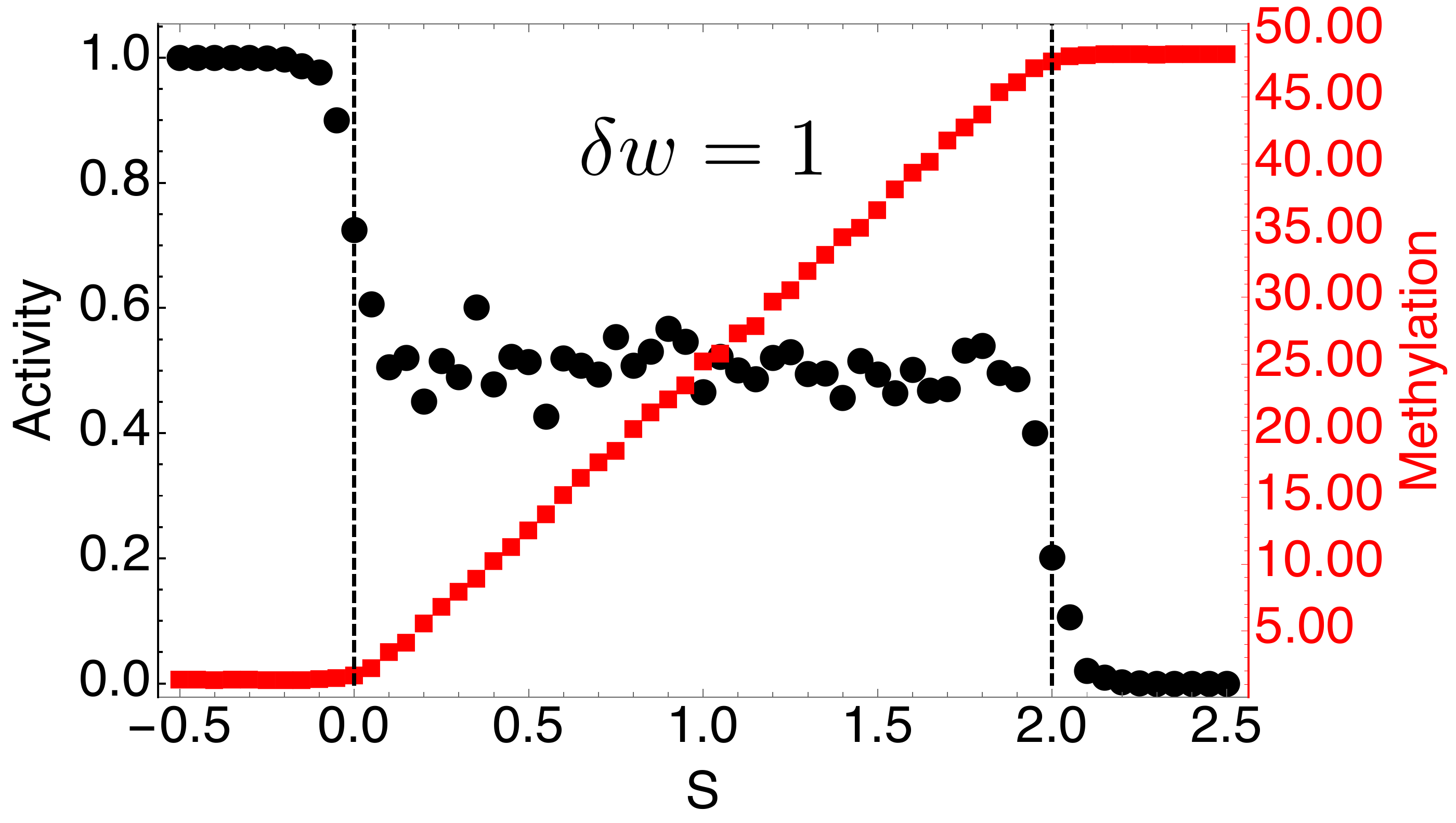}
                 \caption {Steady state behavior of the adaptive network. The parameter $S$ is related to the logarithm of the chemoattractant concentration. The steady state probability density is localized along the horizontal methylation axis whenever $\delta w=1$ and consequently 
                 the average methylation level (orange) tracks the chemoattractant concentration. In this regime, the activity of the network (black) is maintained at a set point over the same wide range of $S$. The maximum number of methylation levels has been set to M=48 here. The numerical results were obtained from networks with quenched disorder in the kinetic rates.}
                                 \label{fig:AdaptMethylation}
 
\end{figure}

In Fig.~\ref{fig:AdaptMethylation} we provide numerical results obtained from simulations with $N=48$ methylation levels~\footnote{We chose $N=48$ methylation levels since previous theoretical and experimental results~\cite{Skoge2011,Skoge2013} suggest that concentration sensing is done by allosterically coupled groups of chemoreceptors. In the SI we present numerical results from simulations with networks containing $10 \leq N \leq 48$. The theoretical predictions impose constraints on fluctuations in all these networks.}. Quenched disorder was introduced in the rates of the kinetic network. The parameter $S$ is a logarithmic function of the chemoattractant concentration. 
We find that the probability density is localized along the methylation axis, $p(m)\sim \exp(-|m-m_0|/\eta)$, where $p(m)$ denotes the probability of observing a methylation level $m$, $m_0$ is determined by the ligand concentration, and $\eta$ is a localization length~\footnote{A similar form for $p(m)$ was derived in~\cite{sartori2015free} for a specific model of chemotaxis adaptation. Our framework generalizes these results.}, whenever $\delta w=1$. The average methylation level tracks the chemoattractant concentration in this regime. 
\begin{figure}[tbp]
\includegraphics[width=0.5\textwidth]{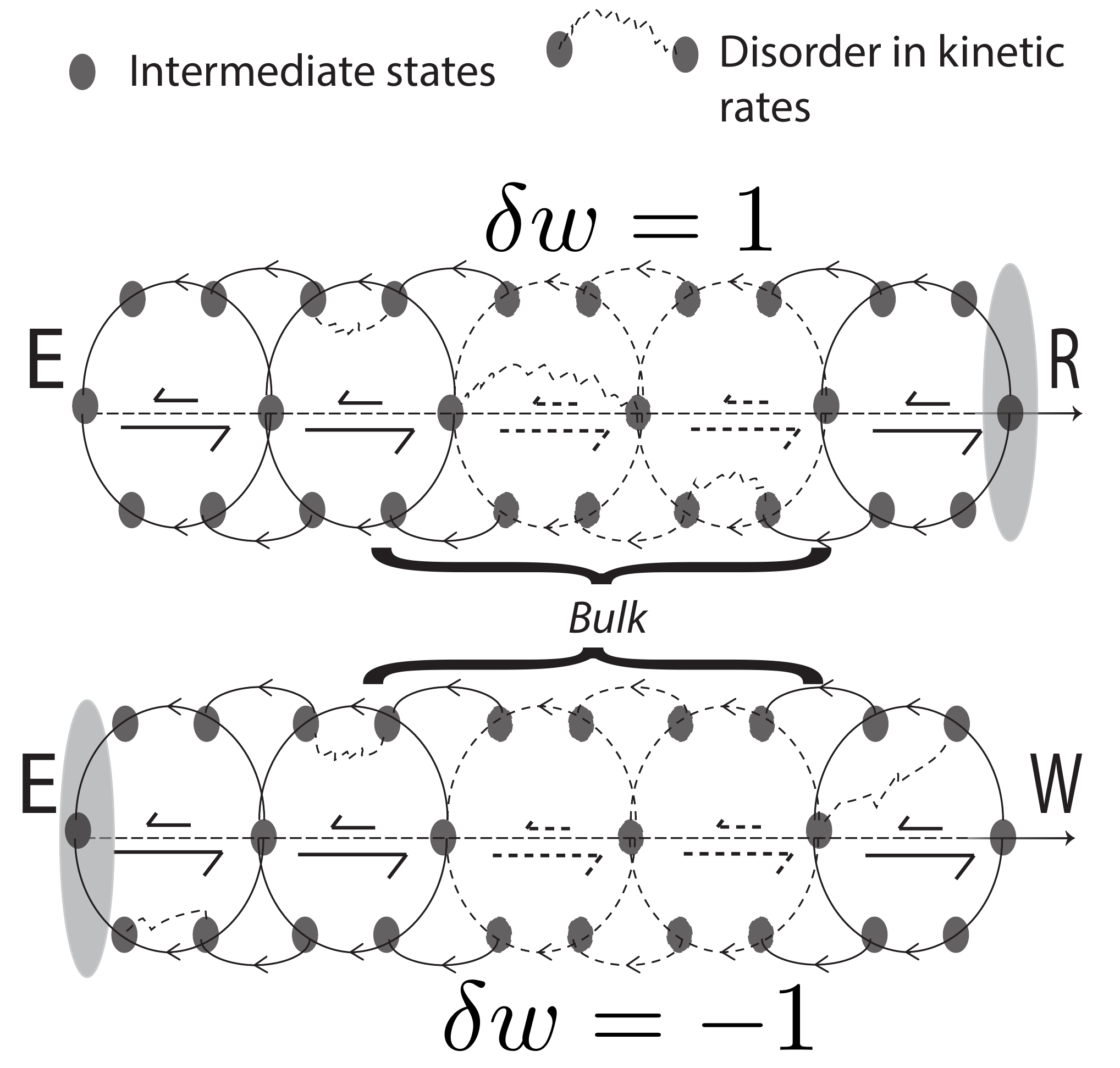}
                                                  \caption{Markov state representation of kinetic proofreading mechanisms. These networks can be viewed as a combination of a periodic bulk phase which terminates at the Right (R)/Wrong (W) products on one end and the reactants ($E$) on the other end. Efficient and robust proofreading can be achieved by tuning the winding numbers of the bulk networks so that the probability distribution is localized at the products end for $R$ and the reactants end for $W$.}
                          \label{fig:kp_example}
                          \end{figure}

A robust adaptive network is defined as one that maintains a set activity of $a=a_0$ for a wide range of chemoattractant concentrations (and other perturbations).  In the SI we show that chemoattractant independent activity is ensured as long as the topological constraint $\delta w=1$ is satisfied and the probability density is localized along the methylation axis as specified above. Further, the response of the system is insensitive to disorder in the kinetic rates in this topologically protected regime (Fig.~\ref{fig:AdaptMethylation} and SI Fig 6)

When the winding number mismatch is $\delta w=-1$, fluctuations in methylation levels are governed by the distribution, $p(m)\sim k_1 \exp(m/\eta_1) + k_2 \exp((M-m)/\eta_2)$ where $k_1$ and $k_2$ are functions of the parameters of the network and $\eta_{1,2}$ are localization lengths. The mean activity is not ligand independent (Fig.~\ref{fig:AdaptMethylation}) in this regime. A topological transition separates the regime exhibiting robust adaptation from the regime in which adaptation is not achieved.

Topologically protected localized modes can promote robustness of kinetic proofreading, a non-equilibrium mechanism that enhances enzymatic specificity~\cite{Hopfield1974}. An enzyme $E$ might be faced with a substrate $R$ that is meant to be processed into a product but is hindered by the presence of a chemically similar undesirable substrate $W$. 
An intuitive way to understand proofreading is through localization (Fig.~\ref{fig:kp_example}); despite substrates $R$ and $W$ having very similar kinetics when binding with an enzyme $E$, reactions with desired substrate $R$ should be localized near products while reactions with $W$ should be localized near reactants \cite{Murugan2012,Murugan2016}. 
Previous literature has investigated many differing models and assumptions on the kinetics of $R$ and $W$ and driving forces that lead to differing localization and hence proofreading \cite{Sartori2013, Murugan2012, Ronde2009,Qian2007}. In contrast, our results on topological protection provide a simple necessary and sufficient condition for efficient proofreading. We view the proofreading network as one bulk phase with one set of kinetics when the enzyme processes $R$ and another set of kinetics when it processes $W$. The products and reactants end of the network correspond to the boundary between the bulk phase and the vacuum. 
The kinetics of $R$ need to be such that the winding number is $w = 1$, localizing it at the products end, while the kinetics of $W$ need to have winding number $w = -1$, localizing it at the reactants end. Unlike the case of adaptation, the localization here is between one bulk phase and the vacuum and not between two bulk phases.

 \section{Concluding remarks}
Our results demonstrate that non-equilibrium systems can support topologically supported localized modes that resemble modes found in topological insulators and meta-materials. These protected modes can provide a general and compact framework to understand the robust functioning of microscopic biochemical mechanisms. 
Specifically, we show that non-equilibrium biochemical models routinely used to study adaptation and kinetic proofreading can support topologically protected modes and can hence function robustly in the presence of disorder.

 \section{acknowledgements}{We gratefully acknowledge conversations with Tom Witten, Aaron Dinner, Mike Rust and particularly Pankaj Mehta.}

%

\end{document}